\documentclass[aps,prl,reprint,preprintnumbers,superscriptaddress,amsmath,bibnotes,longbibliography]{revtex4-2}

\usepackage{microtype}
\usepackage{newtxtext,newtxmath,xspace}
\usepackage{amsbsy,bm,bbold}
\usepackage{siunitx}

\usepackage{graphicx}
\usepackage{dcolumn}
\usepackage[colorlinks,
            linkcolor=blue,
            anchorcolor=blue,
            citecolor=blue,
            urlcolor=blue,
            filecolor=blue,
            menucolor=blue,
            runcolor=blue]{hyperref}
\usepackage[mathlines]{lineno}
\usepackage{soul}
\usepackage{color}
\usepackage{ulem}

\newcommand{\beq}{\begin{equation}}
\newcommand{\eeq}{\end{equation}}
\newcommand{\bea}{\begin{eqnarray}}
\newcommand{\eea}{\end{eqnarray}}

\begin{document}
\title{Emergent ferromagnetic ladder excitations in heavy fermion superconductor CeSb$_{2}$}

\author{Zhaoyang~Shan}
\affiliation{Center for Correlated Matter and School of Physics, Zhejiang University, Hangzhou 310058, China}
\author{Yangjie~Jiao}
\affiliation{Center for Correlated Matter and School of Physics, Zhejiang University, Hangzhou 310058, China}
\author{Jiayu~Guo}
\affiliation{Center for Correlated Matter and School of Physics, Zhejiang University, Hangzhou 310058, China}
\author{Yifan~Wang}
\affiliation{Center for Correlated Matter and School of Physics, Zhejiang University, Hangzhou 310058, China}
\author{Jinyu~Wu}
\affiliation{Center for Correlated Matter and School of Physics, Zhejiang University, Hangzhou 310058, China}
\author{Jiawen~Zhang}
\affiliation{Center for Correlated Matter and School of Physics, Zhejiang University, Hangzhou 310058, China}
\author{Yanan~Zhang}
\affiliation{Center for Correlated Matter and School of Physics, Zhejiang University, Hangzhou 310058, China}
\author{Dajun~Su}
\affiliation{Center for Correlated Matter and School of Physics, Zhejiang University, Hangzhou 310058, China}
\author{Devashibhai~T.~Adroja}
\affiliation{ISIS Facility, STFC Rutherford Appleton Laboratory, Harwell Oxford, Oxfordshire OX11 0QX, United Kingdom}
\affiliation{Highly Correlated Matter Research Group, Physics Department, University of Johannesburg, P.O. Box 524, Auckland Park 2006, South Africa}
\author{Christian~Balz}
\affiliation{ISIS Facility, STFC Rutherford Appleton Laboratory, Harwell Oxford, Oxfordshire OX11 0QX, United Kingdom}
\affiliation{Neutron Scattering Division, Oak Ridge National Laboratory, Oak Ridge, Tennessee 37831, USA}
\author{Matthias~Gutmann}
\affiliation{ISIS Facility, STFC Rutherford Appleton Laboratory, Harwell Oxford, Oxfordshire OX11 0QX, United Kingdom}
\author{Yu~Liu}
\affiliation{Center for Correlated Matter and School of Physics, Zhejiang University, Hangzhou 310058, China}
\author{Huiqiu~Yuan}
\affiliation{Center for Correlated Matter and School of Physics, Zhejiang University, Hangzhou 310058, China}
\affiliation{Collaborative Innovation Center of Advanced Microstructures, Nanjing 210093, China}
\affiliation{State Key Laboratory of Silicon and Advanced Semiconductor Materials, Zhejiang University, Hangzhou 310058, China}
\affiliation{Institute for Advanced Study in Physics, Zhejiang University, Hangzhou 310058, China}
\author{Zhentao~Wang}
\email{ztwang@zju.edu.cn}
\affiliation{Center for Correlated Matter and School of Physics, Zhejiang University, Hangzhou 310058, China}
\author{Yu~Song}
\email{yusong\_phys@zju.edu.cn}
\affiliation{Center for Correlated Matter and School of Physics, Zhejiang University, Hangzhou 310058, China}
\author{Michael~Smidman}
\email{msmidman@zju.edu.cn}
\affiliation{Center for Correlated Matter and School of Physics, Zhejiang University, Hangzhou 310058, China}

\date{\today}

\begin{abstract}
Low-dimensional spin fluctuations play a crucial role in unconventional superconductors, with quasi-one-dimensional spin excitations potentially linked with spin-triplet superconductivity. The heavy fermion superconductor CeSb$_2$  exhibits an unusual large inverted \textit{S}-shaped upper critical field that suggests a possible triplet pairing state within its pressure-induced superconducting dome. Using inelastic neutron scattering, we discover quasi-one-dimensional magnetic excitations in CeSb$_2$ emerging from nearly square Ce layers with minor orthorhombic deformation. We show that the data are well described by a ferromagnetic spin ladder model, where the ``rungs'' of the ladder straddle Ce bilayers. Moreover,  we find that diffuse excitations akin to those in the ordered phase persist well above $T_{\rm N}$, suggesting that quasi-one-dimensional ferromagnetic paramagnons may significantly contribute to the unusual superconductivity that appears under pressure once magnetic order is suppressed.

\end{abstract}

\maketitle

A distinguishing feature of numerous unconventional superconductors, including cuprate, iron-based, organic, and heavy fermion types, is that spin fluctuations act as the primary ``pairing glue'', in contrast to the electron-phonon coupling found in conventional BCS superconductors~\cite{ScalapinoDJ2012_RMP,MiyakeK1986}. The magnetic nature of the pairing mechanism was highlighted through inelastic neutron scattering (INS) studies of the spin excitations in several of these material classes, which reveal a reduction of the magnetic exchange energy in the superconducting state that greatly exceeds the Cooper pair condensation energy~\cite{DemlerE1998,WooH2006,StockC2008,StockertO2011,WangM2013}. Moreover, the close resemblance between the paramagnonic spin excitations crucial for unconventional superconductivity and the spin-wave excitations in nearby magnetically ordered phases suggests that the same magnetic exchange interactions underlie both phenonema~\cite{LipscombeOJ2007,LeTaconM2011,ArndtJ2011,LiuM2012,DeanMPM2013,WangM2013,ZhouKJ2013}. While this connection is widely supported, the precise microscopic description that promotes magnetically driven superconductivity remains to be fully elucidated~\cite{SimethW2023_CeIn3,StockC2015,DasP2014,SongY2021,PasztorovaJ2019,SmidmanM2023_RMP,Chen2024}. In particular, while high-temperature and most heavy fermion superconductors are typically driven by antiferromagnetic fluctuations, ferromagnetism offers a promising route for triplet superconductivity~\cite{FayD1980,SaxenaSS2000,AokiD2001,HuyNT2007}.

Reduced dimensionality often enhances spin fluctuations, which is crucial for unconventional superconductivity, as exemplified by the quasi-two-dimensional high-$T_{\rm c}$ cuprate and iron-based superconductors. In the context of heavy fermion superconductivity, this is demonstrated by the increase of the maximum superconducting $T_{\rm c}$ by an order of magnitude from cubic CeIn$_3$~\cite{MathurND1998} to layered CeRhIn$_5$ and CeCoIn$_5$~\cite{HeggerH2000,PetrovicC2001}. The enhanced quasi-two-dimensionality of CeRhIn$_5$ relative to CeIn$_3$ is clearly reflected in the spin excitation spectra~\cite{KnafoW2003,SimethW2023_CeIn3,DasP2014,StockC2015}, which reveal significantly weaker out-of-plane exchange interactions in  CeRhIn$_5$ compared to those within the basal plane~\cite{DasP2014}. Even more intriguing phenomena arise in one-dimensional systems, whereby several quasi-one-dimensional (q1D) materials are candidate spin-triplet superconductors, whose upper critical fields apparently exceed the Pauli limit along certain crystallographic directions. Examples of such q1D systems include the organic superconductor (TMTSF)$_2$PF$_6$~\cite{LeeIJ1997}, Li$_{0.9}$Mo$_6$O$_{17}$ ~\cite{XuX2009}, and K$_2$Cr$_3$As$_3$~\cite{BaoJK2015,BalakirevFF2015}. In K$_2$Cr$_3$As$_3$ which is formed from Cr$_3$As$_3$ chains, triplet superconductivity is also deduced from the NMR Knight shift~\cite{YangJ2021}, while there is also evidence for antiferromagnetic as well as relatively isotropic ferromagnetic fluctuations~\cite{YangJ2021,TaddeiKM2023_1D}. More recently, UTe$_2$ has emerged as a candidate triplet superconductor with a highly unusual upper critical field with reentrant superconductivity~\cite{RanS2019_science,AokiD2019,KnebelG2019,JiaoL2020}. Its crystal structure consists of U ladders, while INS reveals predominantly incommensurate AFM spin fluctuations, and the degree to which these correspond to low-dimensional ladder like excitations is unresolved~\cite{Xu2019,DuanC2021,DuanC2020,Knafo2021,Butch2022}.

CeSb$_{2}$ is a heavy fermion magnet with a moderately enhanced Sommerfeld coefficient of about 50 mJ~mol$^{-1}$~K$^{-2}$ \cite{Luccas2015,ZhangY2022}. Recently, heavy fermion superconductivity was revealed following the pressure-induced suppression of magnetic order, where there is a dome with a maximum $T_{\rm c}$ of \qty{0.22}{K}~\cite{SquireOP2023}. Unlike typical Ce-based heavy fermions, CeSb$_2$ exhibits potential triplet superconductivity. This is evidenced by an upper critical field that significantly exceeds the Pauli limit and displays an inverted \textit{S}-shaped curve, reminiscent of some U-based materials  \cite{Aoki2009,KnebelG2019,AokiD2022_review}. While CeSb$_2$ has a locally noncentrosymmetric crystal structure, similar to CeRh$_2$As$_2$ in which the upper critical field is also greater than the Pauli limit, unlike the latter there is no evidence of a field-induced transition from an even-parity to an odd-parity state \cite{Khim2021,Landaeta2022,Kibune2022}. Alternatively, it was suggested that the robustness of superconductivity in-field may arise from the pairing of strongly coupled ultraheavy quasiparticles \cite{SquireOP2023}.

At ambient pressure, CeSb$_{2}$ exhibits multiple magnetic phase transitions \cite{TrainerC2021,Zhang2017},
whereby it orders magnetically below 15.5~K, with successive transitions at 11.5, 9.5 and 6.5~K. While the nature of the magnetic phases remains unclear, it is suggested that the transitions below  11.5 and 9.5~K are to different antiferromagnetic states, while there is evidence for  a ferromagnetic component below 6.5~K. Magnetization measurements show that the easy direction lies within the $ab$ plane \cite{BudkoSL1998}. On the other hand, the only magnetic Bragg peaks observed in neutron diffraction correspond to a unidirectional in-plane propagation vector $\bm{k} =(\pm1/6, -1 ,0)$ that appear below 9.8 K~\cite{LiuB2020}.

\begin{figure}[tb]
\begin{center}
\includegraphics[width=0.71\columnwidth]{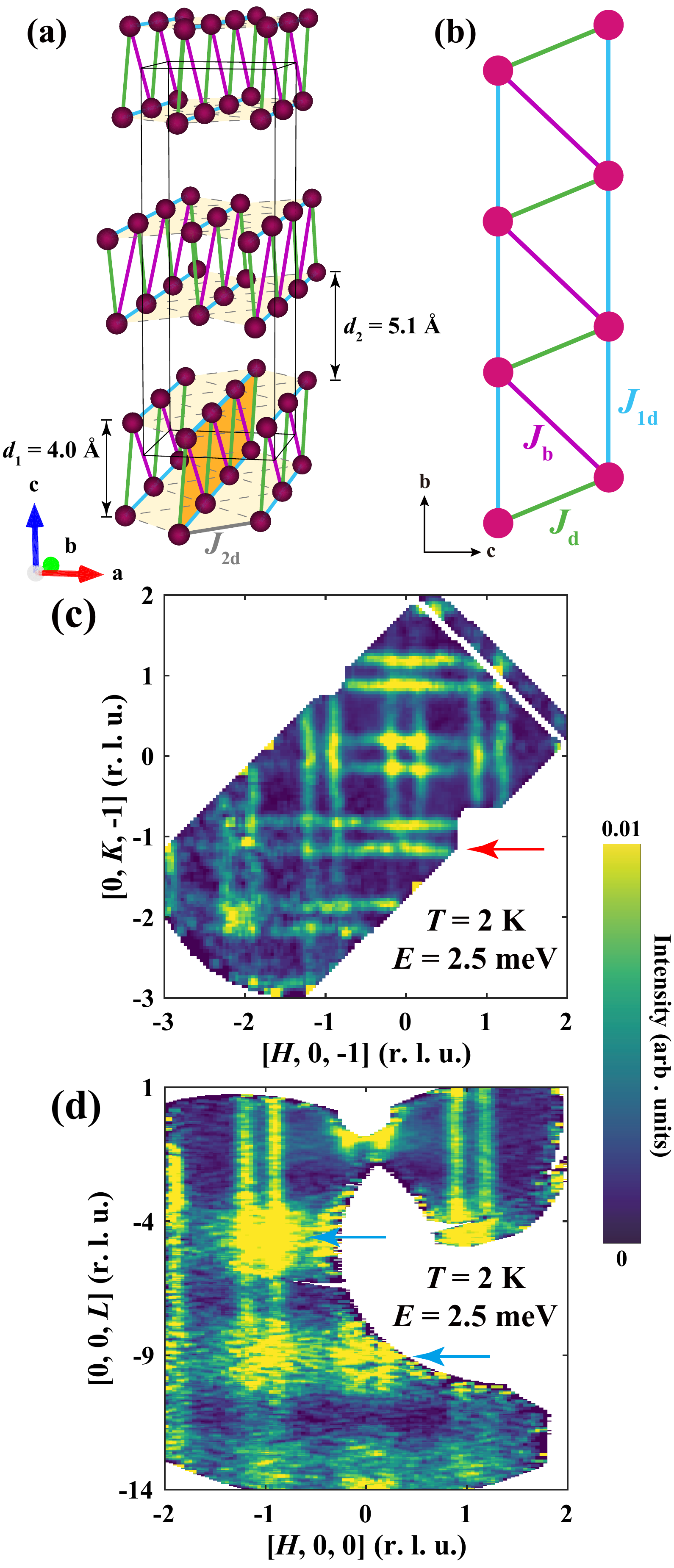}
\end{center}
\caption{(a) Arrangement of Ce atoms in the crystal structure of CeSb$_2$, consisting of rectangular layers staggered along the $c$~axis. Note that six layers are displayed which correspond to three bilayers with intrabilayer separation of $d_1$ and interbilayer distance $d_2$, and the black solid lines highlight the unit cell.  (b) Schematic of the spin ladder structure within the bilayers. The magnetic exchange interactions considered for the spin-ladder model described in the text are also labelled in (a) and (b). Constant energy slices of the INS intensity for $E=2.5\pm\qty{0.5}{meV}$ at \qty{2}{K} are displayed for the (c) $[H,K,-1]$, and (d) $[H,0,L]$  planes. The colors correspond to the intensity in arbitrary units. The red arrow in panel (c) highlights the rodlike quasi-one dimensional excitations, while the blue arrows in (d) show the maxima of the amplitude modulation at $L=-4.5$ and $-9$.}
\label{fig1}
\end{figure}

CeSb$_{2}$ crystallizes in an orthorhombic structure with space group $Cmce$, with lattice parameters $a=6.28$, $b=6.13$ and $c=18.24$~{\AA}. The similar in-plane lattice parameters suggest a propensity towards twinning of orthorhombic domains in crystalline samples. The layered arrangement of the Ce atoms is shown in Fig.~\ref{fig1}(a), where each unit cell consists of four rectangular staggered Ce layers along the $c$ axis, which form a bilayer structure with an intrabilayer separation of $d_1=4.0$~{\AA} and interbilayer distance of  $d_2=5.1$~{\AA}. Conversely, there are only small differences between the Ce-Ce distances along the $a$ and $b$  axes. Remarkably, while this structure might suggest a two-dimensional or bilayer nature of the magnetic excitations, in this Letter we show using INS that they in fact have distinctively q1D characteristics that can be captured by a ferromagnetic spin ladder model. Importantly, q1D paramagnons persist in the paramagnetic state well above $T_{\rm N}$, suggesting that these could be important for realizing potential spin-triplet superconductivity under pressure.

Single crystals of CeSb$_{2}$ were grown by a Sb self-flux method~\cite{ZhangS2021}. Samples were checked using the SXD diffractometer \cite{Keen2006}, and INS was performed on $\sim \qty{5}{g}$ of selected coaligned single crystals using the LET spectrometer \cite{Bewley2011} at ISIS with [110] and [001] in the horizontal scattering plane, operating in the multiple-$E_{i}$ mode, with incident energies $E_{i}=15$, 6, 3 and \qty{2}{meV}. Elastic scattering confirms the orthorhombic twins observed previously~\cite{LiuB2020,SI}. 
\begin{figure*}[tb]
\begin{center}
\includegraphics[width=0.95\textwidth]{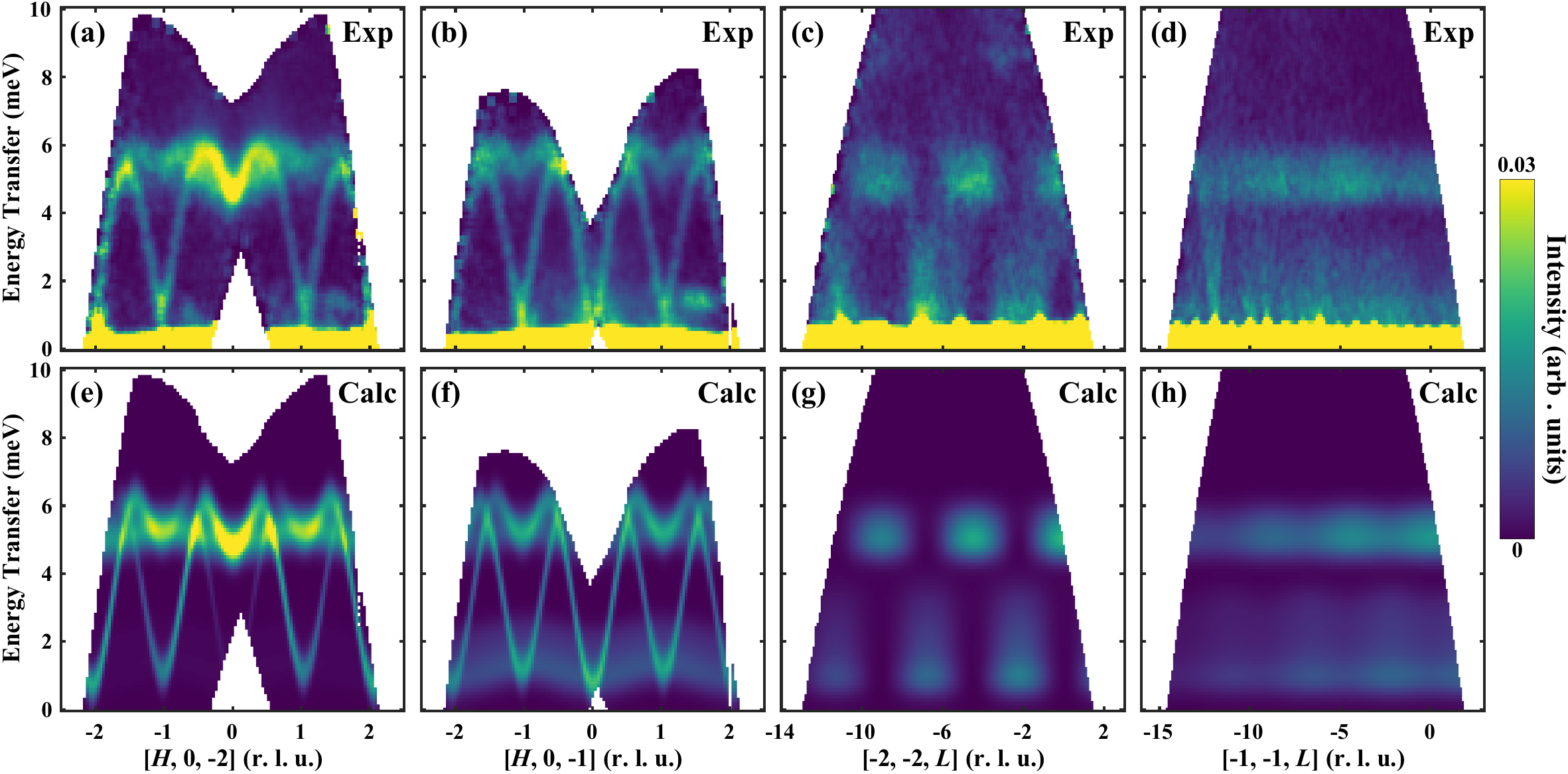}
\end{center}
\caption{Color plots of the INS intensity (in arbitrary units)  at \qty{2}{K} along the (a) $[H, 0, -2]$, (b) $[H, 0, -1]$, (c) $[-2, -2, L]$ and (d) $[-1, -1, L]$ directions. The momenta were integrated over $\pm0.2$ r.l.u.. Panels (e)-(h) show LSW calculations of the INS intensity using the spin-ladder model described in the text for the respective directions in (a)–(d), with model parameters $J_\text{b}=\qty{-2.2}{meV}$, $J_\text{d}=\qty{-1.8}{meV}$, $J_\text{1d}=\qty{-1.4}{meV}$, $J_\text{2d}=\qty{-0.1}{meV}$, and $\delta=1.2$.}
\label{fig2}
\end{figure*}

The q1D nature of the spin excitations is demonstrated by the constant energy slices centered around \qty{2.5}{meV} in Figs.~\ref{fig1}(c) and (d)  measured at \qty{2}{K} (below $T_{\rm N}$) in the $[H,K,-1]$ and $[H,0,L]$ planes, respectively. For magnons propagating within the Ce-layers, the scattering in the $[H,K,-1]$ plane would be expected to exhibit ring-like features, but such features are not observed at any energies. Instead, it can be seen that there are rod-like features in the scattering plane (see the red arrow in Fig.~\ref{fig1}(c)), showing that the magnons are constrained along one of the in-plane directions, as observed in spin-chain magnetic oxides~\cite{JohnstoneGE2012,GaoS2024}. Similar behaviors persist in energy slices integrated across different energy ranges, up to the top of the magnon band at around \qty{5}{meV}~\cite{SI}. Note that the presence of rods oriented along the two perpendicular directions is a consequence of the crystal twinning found in elastic scattering \cite{SI}.

Similarly, the $[H,0,L]$ plane in Fig.~\ref{fig1}(d) also exhibits rodlike scattering along the $L$ direction, showing that there is no long-range interlayer propagation of magnons. Put together with the observations in the $[H,K,-1]$ plane, this demonstrates a one-dimensional nature of the spin excitations in CeSb$_2$. On the other hand, although there is a lack of dispersion in the out-of-plane direction, it can be seen that there is a modulation of the excitation intensity, with maxima around $L=-4.5$ and $L=-9$ [see the blue arrows in Fig.~\ref{fig1}(d)]. Such amplitude modulations are characteristic of in- and out-of-phase bilayer excitations~\cite{PailhesS2003,HaydenSM1996,Xie2018}, indicating that there is a strong coupling within the Ce bilayers, but interbilayer couplings are extremely weak, and therefore the in-plane q1D excitations correspond to ladder excitations.

The dispersion of the magnon excitations along one of the in-plane directions is shown in Figs.~\ref{fig2}(a) and (b), for even and odd $L$, respectively.  There are two distinct well-defined branches, an acoustic branch with a small spin-gap of around $\Delta\approx \qty{1}{meV}$, and a weakly dispersing optical mode, whereby both branches have a band top at around \qty{6}{meV}. Note that for $L=-2$, the acoustic mode has very weak intensity near $H=0$. As both these modes stem from $\bm{k}=0$, this strongly suggests a predominantly ferromagnetic nature of the excitations. In the out-of-plane directions shown in Figs.~\ref{fig2}(c) and (d), there is a lack of dispersion of the excitations which are also more diffuse. Along $[-2,-2,L]$, the aforementioned modulation of the amplitudes is also observed, while along $[-1,-1,L]$ such features are much less pronounced.

In the crystalline-electric field of CeSb$_2$, the $J=5/2$ multiplet of Ce$^{3+}$ is split into three Kramers doublets, and the ground state doublet corresponds to an effective $S=1/2$ system. Therefore, to capture the characteristic q1D features of these data, we consider a $S=1/2$ spin-ladder model for  the magnetic exchange interactions~\cite{UhrigGS2005,VuleticT2006,EcclestonRS1998,WuS2020,TakahashiH2015}:
\begin{equation}
\mathcal{H}  = \sum_{\langle ij \rangle} J_{ij}(S_{i}^{x}S_{j}^{x} + \delta S_{i}^{y}S_{j}^{y} + S_{i}^{z}S_{j}^{z}).
\label{eq:Ham}
\end{equation}
\noindent Four magnetic exchange parameters $J_{ij}$ were included to construct a minimal model, whose exchange pathways are displayed in Figs.~\ref{fig1}(a) and (b). Figures~\ref{fig2}(e)-(h) show the calculated INS spectra performed within the framework of linear spin wave (LSW) theory using the \textsc{SpinW} software~\cite{TothS2015}, incorporating the Ce$^{3+}$ form factor \cite{Blume1962}. Because of the crystallographic twinning, the spectra for two perpendicular directions were summed with equal weights. So as to compare with the data, the calculated LSW spectra are plotted with bin windows and integration widths matching those of the data cuts.

We start by considering the observed intensity modulation along the $L$ directions, which necessitates the inclusion of intra-bilayer couplings $J_{\rm d}$ and $J_{\rm b}$, with respective bond lengths of 4.33 and  6.01 $\AA$, while interbilayer coupling is negligible. We find that ferromagnetic coupling of the bilayers with $J_{\rm b}={-2.2}$ and $J_{\rm d}=\qty{-1.8}{meV}$ effectively reproduces the periodicity of the magnetic excitations along the $L$ direction, and therefore the ``rungs'' of the ladders are predominantly ferromagnetic interactions.

\begin{figure}[tb]
\begin{center}
\includegraphics[width=\columnwidth]{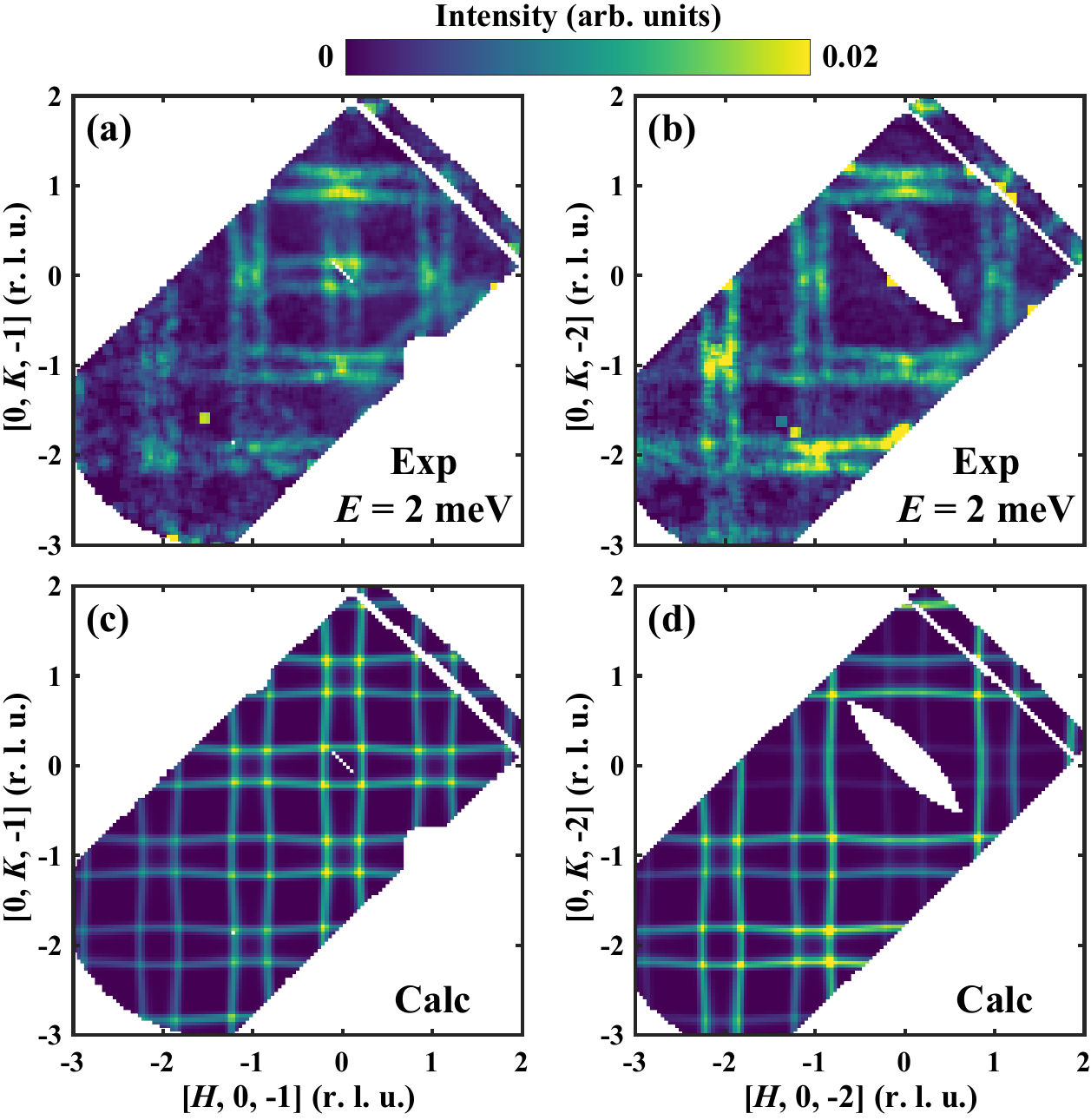}
\end{center}
\caption{INS intensity at \qty{2}{K}  in the $[H,K]$ plane for (a)  $L = -1$ and, (b) $L = -2$. The energy was integrated over a range $2\pm \qty{0.3}{meV}$. (c) and (d) display the corresponding intensities calculated by LSW using the spin-ladder model described in the text, with model parameters $J_\text{b}=\qty{-2.2}{meV}$, $J_\text{d}=\qty{-1.8}{meV}$, $J_\text{1d}=\qty{-1.4}{meV}$, $J_\text{2d}=\qty{-0.1}{meV}$, and $\delta=1.2$.}
 \label{fig3}
\end{figure}

Within the $ab$~plane, the nearest neighbor exchange path $J_{\rm 2d}$, with a bond length of 4.39~$\AA$, connects each Ce atom to four others along the diagonals, and so significant coupling would give rise to a 2D rather than 1D exchange path. Therefore this coupling must be small, corresponding to interactions between ladders. As shown in Figs.~\ref{fig3}(a) and (b), the rods are not perfectly straight but exhibit a slight bending. Although there are multiple possible exchange pathways coupling the ladders, the bending can be effectively accounted for with a small value of $J_{\rm 2d}=\qty{-0.1}{meV}$, as shown in Figs.~\ref{fig3}(c) and (d). On the other hand, to realize strongly dispersing magnons along one of the in-plane directions, a ferromagnetic interaction $J_{\rm 1d}=\qty{-1.4}{meV}$ along the in-plane direction with the shorter Ce-Ce distance (6.13 $\AA$ along the $b$ axis) is also included. Lastly, as the magnetization measurements show that the easy axis lies within the $ab$~plane~\cite{BudkoSL1998}, a uniaxial exchange anisotropy $\delta>0$ is included in Eq.~\eqref{eq:Ham}. We find that a value of $\delta=1.2$ reproduces the \qty{1}{meV} spin-gap of the acoustic magnon mode.

With this set of parameters, it can be seen in Fig.~\ref{fig2} as well as in the Supplemental Material~\cite{SI} that the observed magnetic excitations can be reproduced, including the modulations in the intensity. Furthermore, it can be seen in Fig.~\ref{fig3} that the spin-ladder model can well reproduce the main features of the constant energy slices, confirming the q1D nature of the magnetic excitations, while the slight bending of the rods is accounted for by the small interladder coupling  $J_{\rm 2d}$.

\begin{figure}[tb]
\begin{center}
\includegraphics[width=0.935\columnwidth]{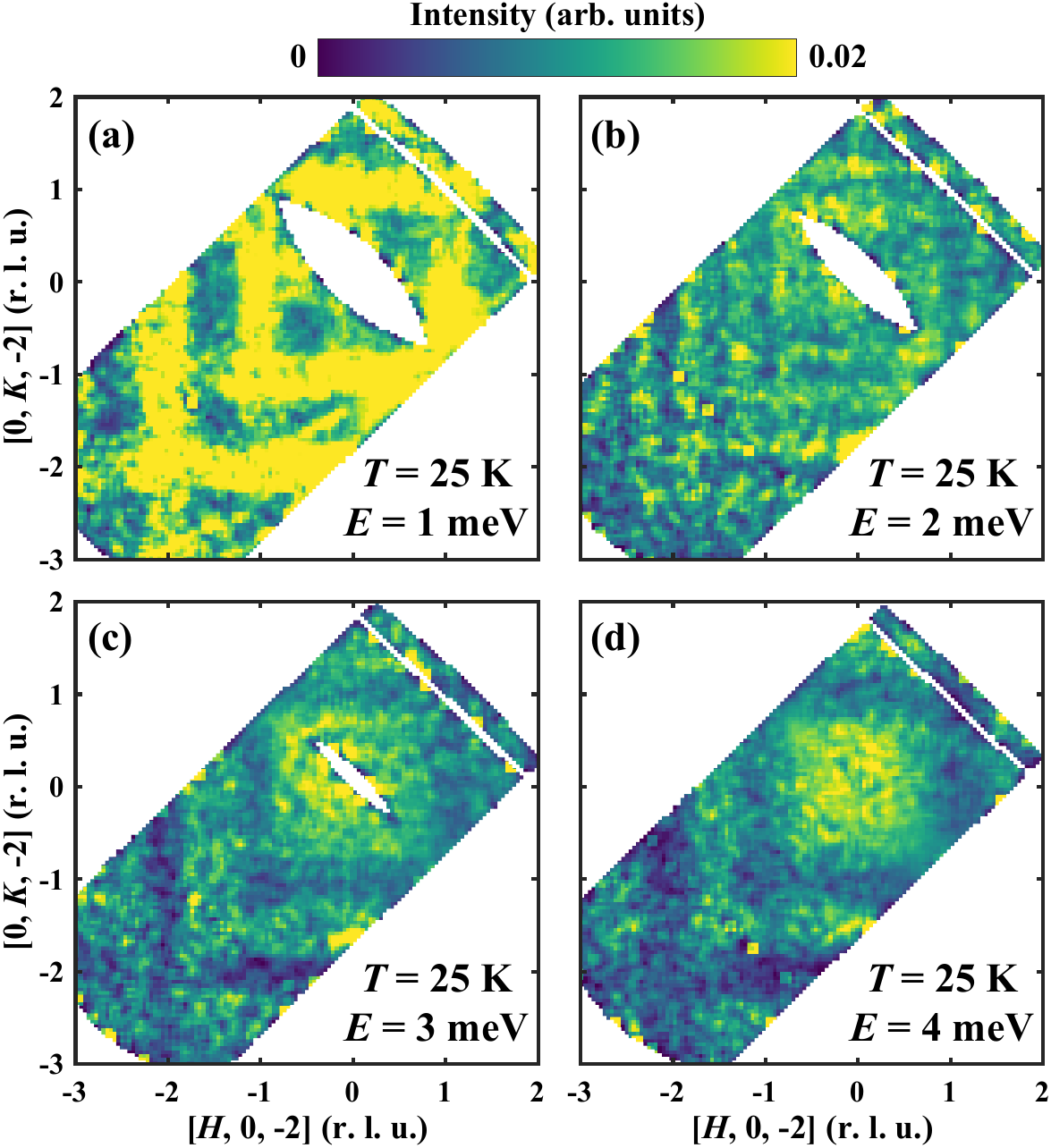}
\end{center}
\caption{Constant energy slices of the CeSb$_{2}$ INS spectra measured in the paramagnetic state at $T = \qty{25}{K}$ in the $[H, K, -2]$ plane at four different neutron energy transfers $E\pm0.3~$meV of: (a) 1, (b) 2, (c) 3, and (d) $\qty{4}{meV}$.}
 \label{fig4}
\end{figure}

Note that  this simplified model has a $\bm{k}=0$ magnetic ground state, while previous neutron diffraction measurements~\cite{LiuB2020}, as well elastic cuts of our data~\cite{SI}, reveal a complex magnetic ground state with $\bm{k} = (\pm1/6, -1 ,0) $. While the detailed magnetic structures of the different phases of CeSb$_2$ are yet to be determined, we note that a magnetic ground state with such a long range in-plane periodicity could be realized by inclusion of additional frustrated nearest-neighbor and next-nearest neighbor interactions either parallel or perpendicular to the ladder direction, in analogy to various well-studied Ising spin models~\cite{SelkeW1988_review}, without significantly changing the magnetic excitations that are dominated by the interactions considered in the discussed spin ladder model. Moreover, CeSb$_{2}$ is reported to  undergo a structural transition under pressure, which is suggested to be to the YbSb$_{2}$-type structure \cite{SquireOP2023,Kagayama2000}. In such a structure, the two couplings forming the ``rungs'' of the ladder ($J_{\rm d}$ and $J_{\rm b}$) become equivalent, whereas in our model these parameters have only a slight difference. The orthorhombic twinning of CeSb$_{2}$ samples may also suggest a high-temperature structural instability at ambient pressure, but no signatures of any structural phase transition are reported.

Another characteristic feature of low-dimensional magnetism is the persistence of short-range magnetic correlations  in the paramagnetic state. Figure~\ref{fig4} displays constant energy cuts of INS measurements in the $[H,K,-2]$ plane performed in the paramagnetic state at \qty{25}{K}, well above $T_{\rm N}=\qty{16}{K}$. It is evident that the rodlike excitations persist in the paramagnetic state, and although they become more diffuse, they strongly resemble the low temperature spin-wave excitations, indicating the presence of q1D paramagnons above $T_{\rm N}$. Such magnonlike features in the paramagnetic state of our spin-ladder model have been confirmed by finite-temperature calculations of spin excitations~\cite{SI}.

The presence of q1D paramagnons up to at least $1.5T_{\rm N}$ not only gives further evidence for q1D magnetism, but also has implications for the nature of the unconventional superconductivity. In magnetically driven superconductors, the same magnetic interactions likely give rise to both the magnetic order and superconductivity that are proximate in the phase diagram. Such a link is highlighted in heavy fermion CeCu$_2$Si$_2$ where the dispersive paramagnon mode in superconducting samples that becomes gapped out in the superconducting state~\cite{ArndtJ2011,StockertO2011} is centered precisely at the incommensurate  ordering wavevector of the antiferromagnetically ordered samples~\cite{StockertO2004}. The inverted \textit{S}-shaped upper critical field of CeSb$_{2}$ deviates from typical centrosymmetric Ce-based heavy fermion superconductors. This anomaly signals the possibility of spin-triplet superconductivity, a phenomenon that is more commonly observed in U-based materials. Our results suggest that ferromagnetic q1D fluctuations, similar to those observed in this study, may be responsible for the unusual heavy fermion superconductivity in CeSb$_{2}$.

In summary, we find that a highly unusual q1D magnetism in the heavy fermion superconductor CeSb$_2$ emerges in a layered crystal structure. The magnetic excitations are well accounted for by a model of ferromagnetic ladders running in the plane of the bilayers. These findings highlight CeSb$_2$ as a model system for the realization of q1D magnetism in Kondo lattices as well as other correlated $f$-electron intermetallic systems, and more broadly, for studying the connection between q1D ferromagnetic excitations and unconventional, possibly spin-triplet, superconductivity. Given that photoemission spectroscopy measurements suggest a quasi-two-dimensional electronic structure \cite{ZhangY2022}, it is of particular interest to understand the origin of the q1D ferromagnetic spin-ladder excitations, by developing microscopic models to obtain a unified understanding of the magnetic ground state and interactions.

\begin{acknowledgements}

This work was supported by the National Key R$\&$D Program of China (Grants No.~2022YFA1402200, No.~2023YFA1406303 and No.~2024YFA1408303), the Key R$\&$D Program of Zhejiang Province, China (Grant No.~2021C01002), the National Natural Science Foundation of China (Grant No.~12222410, No.~12174332, No.~12034017, No.~12374124, No.~12274363, and No.~12350710785), the Fundamental Research Funds for the Central Universities (Grant No.~226-2024-00068). Experiments at the ISIS Pulsed Neutron and Muon Source were supported by a beamtime allocation from the Science and Technology Facilities Council (RB2320153)~\cite{LETdata}.

\end{acknowledgements}

\end{document}